\documentclass[12pt,thmsa]{article}
\usepackage{amssymb}

\usepackage{sw20lart}

\input{tcilatex}
\begin{document}

\title{Towards a realistic interpretation of quantum mechanics providing a model of
the physical world}
\author{Emilio Santos \\
Departamento de F\'{i}sica. Universidad de Cantabria. Santander.Spain}
\maketitle

\begin{abstract}
It is argued that a realistic interpretation of quantum mechanics is
possible and useful. Current interpretations, from ``Copenhagen'' to ``many
worlds'' are critically revisited. The difficulties for intuitive models of
quantum physics are pointed out and possible solutions proposed. In
particular the existence of discrete states, the quantum jumps, the alleged
lack of objective properties, measurement theory, the probabilistic
character of quantum physics, the wave-particle duality and the Bell
inequalities are analyzed. The sketch of a realistic picture of the quantum
world is presented. It rests upon the assumption that quantum mechanics is a
stochastic theory whose randomness derives from the existence of vacuum
fields. They correspond to the vacuum fluctuations of quantum field theory,
but taken as real rather than virtual.
\end{abstract}

\section{ Introduction}

Quantum mechanics is extremely efficient for the prediction of experimental
results. In contrast, the interpretation of the quantum formalism has been
the subject of continuous debate since the very begining of the theory\cite
{Jammer}, \cite{WZurek}, \cite{Fine}, \cite{Auletta}, \cite{Laloe} and it
lasts until today\cite{Mittelstaedt1}, \cite{d'Espagnat}, \cite{Beneduci}.
Is there a real problem? Feynman believed that a problem exists when he
stated: ``Nobody understands quantum mechanics''\cite{Feynman}, and many
people agree with that opinion. The fact is that none of the different
interpretations proposed till now offers a clear intuitive picture of the
quantum world. Nevertheless most physicists do not worry for the lack of a
picture and embrace a \textit{pragmatic approach, }good enough in practice.
In contrast with this attitude, in this paper a \textit{realistic
interpretation} of quantum theory is supported, the difficulties for that
interpretation are analyzed and a picture of the quantum world is proposed.

The plan of the paper is as follows. In this Introduction section, a few
general comments are made on two opposite approaches to the interpretation
of quantum mechanics, namely pragmatic and realistic. After that the sketch
is presented of a picture of the quantum world. In section 2 a number of
empirical facts of the quantum domain are analyzed in order to show that
none of them prevents the existence of an intuitive picture of the
microworld. In section 3 the most popular interpretations proposed till now,
from `Copenhagen' to `many worlds', are revisited critically. In section 4
it is discussed the `ensemble interpretation', that supports an
epistemological rather than ontological treatment of the wave function. Also
the closely related subjects of hidden variables models and Bell's
inequalities are commented. Finally in section 5 the proposed picture of the
quantum world is discussed in some detail.

\subsection{The pragmatic approach to quantum mechanics}

Many physicists, not too interested in foundations, accept\textit{\ }a
minimal interpretational framework with the following key features\cite
{Isham}:

1. Quantum theory is viewed as a scheme for predicting the probabilistic
distribution of outcomes of measurements made on suitable prepared copies of
a system.

2. The probabilities are interpreted in a statistical way as referring to
relative frequencies.

In general a physical theory has at least two components\cite{Suppe}: (1)
the formalism, or mathematical apparatus, of the theory, and (2) the rules
of correspondence that establish a link between the formalism and the
results of measurements. For instance, the standard formalism of quantum
mechanics is based on the mathematical theory of Hilbert spaces. In it there
are two essential kinds of operators, density operators, $\hat{\rho},$ that
represent states, and self-adjoint operators, $\hat{A}$, that represent
observables. The link with the measurement results is given by the Born rule
where the `expectation value', $Tr\left( \hat{\rho}\hat{A}\right) ,$ is
assumed to correspond to the statistical mean of the values obtained when
one realizes several measurements on identically prepared systems (which
determines $\hat{\rho})$ by means of an apparatus which corresponds to $\hat{%
A}.$ If we assume that the formalism and the correspondence rules are the
only objects required to define a physical theory, in the sense that the
statistical regularities need not be further explained, then we get what has
been called a \textit{minimal instrumentalistic interpretation} of the theory%
\cite{Redhead}. It might be named \textit{pragmatic approach} or even
qualified as rejection of any interpretation\cite{Peres}.

Most people claiming to support that approach accept the following positions
which go beyond the purely pragmatic attitude:

1. The notion of an individual physical system `having' or `possessing'
values for all its physical quantities is \textit{inappropriate} in the
context of quantum theory.

2. The concept of `measurement' is fundamental in the sense that the scope
of quantum theory is \textit{intrinsecally} restricted to predicting the
results of measurements.

3. The spread in the results of measurements on identically prepared systems
must not be interpreted as reflecting a `lack of knowledge' of some
objectively existing state of affears.

Actually these propositions define an interpretation that has been also
called \textit{instrumentalistic}\cite{Isham}. It is quite different from,
even opposite to, the \emph{realistic} view traditional of classical
physics. Between these two extremes there are a variety of approaches.

\subsection{Realistic interpretations}

In this paper it is supported a realistic interpretation that demands
physical models for the quantum phenomena. This position is not new, it has
been advocated by many people including Einstein as the most distinguished
author. Indeed the celebrated article by Einstein, Podolsky and Rosen\cite
{EPR} (EPR) begins: ``Any serious consideration of a physical theory must
take into account the distinction between the objective reality, which is
independent of any theory, and the physical concepts with which the theory
operates. These concepts are intended to correspond with \emph{the objective
reality}, and by means of these concepts \emph{we picture this reality to
ourselves'' }(my emphasis).

It is true that in the 80 years since the EPR paper the concept of
``objective reality'' has been questioned as not clear. Due to the
difficulties with the interpretation of quantum mechanics, many people
working on foundations dismiss the `realism' of EPR as `naive'. Thus more
sophisticated forms or realism have been proposed\cite{d'Espagnat1}, \cite
{d'Espagnat2}. In any case a discussion about the philosophical aspects of
reality or realism is outside the scope of this paper.

The starting point of this article is the claim that any physical theory
should contain a \emph{physical model} in addition to \emph{the formalism
and rules for the connection with the experiments. }The latter are obviously
essential because they are required for the comparison of the theory with
empirical evidence, which is \emph{the test} for the validity of the theory.
But in my opinion physical models are also necessary in order to \emph{reach
a coherent picture }of\emph{\ }the world. Many quantum physicists apparently
support the useless of pictures, but it is the case that when they attempt
popular explanations of quantum mechanics they propose models, and in fact
rather bizarre ones. For instance it is claimed that quantum mechanics \emph{%
compel us } to believe that there are a multiplicity of ``me'' in parallel
universes or that an atom may be present in two distant places at the same
time. For me this is an indication that the need of ``picture the reality to
ourselves''\cite{EPR} cannot be easily rejected. Furthermore the existence
of physical models might open the possibility for new developments and
applications of quantum theory and therefore it is not a purely academic
question.

It is interesting the contrast between the two great theories of the 20th
century, quantum mechanics and relativity. The latter provides a beautiful
physical model: There is a four-dimensional manifold with intrinsic
curvature and all material objects (e. g. particles or fields) are defined
in that continuum. This is fundamental even for the formulation of quantum
(field) theory. But the calculational tool of general relativity (derived
from the Riemann geometry) is rather involved, the fundamental (Einstein)
equations being nonlinear. In quantum mechanics there is a relatively simple
formalism involving vectors and operators in a Hilbert space. Indeed the
fundamental (Schr\"{o}dinger) equation is linear. However there is no
coherent physical model behind it. I would say that general relativity has 
\emph{physical beauty}, the quantum formalism possesses \emph{mathematical
elegance.}

Historically the renounciation to physical models in quantum mechanics was a
consequence of frustration caused by the failure of the models proposed
during the first quarter of the 20th century. This was specially the case
after Bohr's atom, consisting of point electrons moving in circular orbits
around the nucleus. The model, generalized with the inclusion of elliptical
orbits, produced some progress in the decade after 1913. However it was
increasingly clear that the model was untenable. In 1926 an alternative
model was proposed by Schr\"{o}dinger, who interpreted his wave mechanics as
showing that electrons are continuous charge distributions. As is well known
that model was soon abandoned after the correct criticisms by Bohr,
Heisenberg and other people. Independently Heisenberg had proposed a
formalims, with the name of quantum mechanics, that \emph{explicitly
rejected any model}. Indeed he supported the view that the absence of a
picture was a progress towards a more refined form of scientific knowledge.
The success of the new quantum mechanics in the quantitative interpretation
of experiments, toghether with the failure to find a good physical model of
the microworld, led to the almost universal acceptance of the current view
that models are unnecessary or even misleading.

I do not agree with that wisdom, and this paper is a defence of a realistic
interpretation of the quantum phenomena. I am aware that the task is
extremely difficult, but convinced that many of the obstacles derive from
assumptions unnecessary for the interpretation of the experiments. These
assumptions have been included along the time and are now a part of the
common wisdom. Pointing out the main obstacles and how they might be removed
is the main purpose of this article. It does not pretend to be a coherent
and complete realistic interpretation.

\subsection{A note on epistemology of physics}

In order to make science some previous philosophical questions should be
answered. For instance, what is science?, or what are the criteria to
distinguish science from nonscientific knowledge? I accept the definition of
Karl Popper\cite{Popperlogic}: ``A claim is scientific if it may be refuted
by observations or experiments''. This definition is a consequence of a well
known fact, that is \emph{the possible existence of several different
theories all of them predicting correctly the results of experiments in a
given domain. }In other words the correctness of a theory is sufficient, 
\emph{but not necessary}, for the appropriate prediction of the empirical
facts. For this reason a single experiment may refute a theory but a theory
can never be fully confirmed empirically, and this is essentially the Popper
thesis. As a consequence several different theories may exist able to
predict correctly the empirical results, but suggesting quite different
pictures of the microworld.

Popper\'{}s criterion is good enough as a matter of principle, but it is not
so good in practice. In fact it is the case that rarely an established
theory breaks down as a consequence of a single experiment contradicting it.
As Lakatos\cite{Lakatos} has correctly pointed out, \emph{well tested
theories are protected} in the sense that the empirical refutation of a
single prediction may be interpreted without rejecting the theory, for
instance assuming that the particular model used to analyze the experiment
was too coarse. Indeed it is a historical fact that established theories are
only abandoned, or better superseeded, when there is a new theory in
agreement with the former one in its domain of validity, but possessing a
wider domain or other virtues.

Quantum mechanics is today a fully established theory and therefore it is
very well protected in the sense of Lakatos. I do not only mean protection
in the domain where the theory has been tested. What I want to stress is
that along the years people has introduced a number of assumptions, today
widely accepted, that are additions beyond any possible empirical test.
These unnecessary additions are also protected and, in my opinion, they are
the main cause of the strong difficulties in reaching a realistic physical
model of the quantum world.

Most working quantum physicists adhere to the pragmatic approach as
commented above. The support has its roots in a ``positivistic'' attitude.
Positivism is the philosophical doctrine that, in a broad sense, states that
all knowledge should be founded on empirical evidence. If the statement is
applied to scientific knowledge, it is accepted by everybody. But in a more
strict sense it is a tendency to give value to the empirical data in
detriment of the theoretical elaborations. For instance this was the opinion
of Ernst Mach, who rejected the concept of atom because at that time (around
1900) atoms had not been directly observed.

Positivism was also behind Heisenberg initial formulation of quantum
mechanics resting upon the belief that only sets of numbers corresponding to
the possible results of measurements should enter the theory. This led him
to elaborate quantum mechanics as a calculational tool involving matrices
(that was sometimes called `matrix mechanics'.) The combination of
mathematical formalism and empirical results almost without further
theoretical elaboration permeates the interpretation of quantum mechanics
till now. A clear confrontation between the positivistic and realistic
epistemologies appears in the conversation of Heisenberg with Einstein that
took place in Berlin 1926, as remembered by Heisenberg himself\cite
{Heisenberg26}. The most relevant part is reproduced in the following.

``Einstein opened the conversation with a question that bore on the
philosophical background of my recent work.`What you have told us sounds
extremely strange. You assume the existence of electrons inside the atom,
and you are probably quite right to do so. But you refuse to consider their
orbits, even though we can observe electron tracks in a cloudchamber. I
should very much like to hear more about your reasons for making such
strange assumptions'. `We cannot observe electron orbits inside the atom', I
must have repIied, `but the radiation which an atom emits during discharges
enables us to deduce the frequencies and corresponding amplitudes of its
electrons. After all, even in the older physics wave numbers and amplitudes
could be considered substitutes for electron orbits. Now, since \textit{a
good theory must be based on directly observable magnitudes}, I thought it
more fitting to restrict myself to these, treating them, as it were, as
representatives of the electron orbits.' `But you don't seriously believe',
Einstein protested, `that none but observable magnitudes must go into a
physical theory?'. `Isn't that precisely what you have done with
relativity?' I asked in some surprise. `After all, you did stress the fact
that it is impermissible to speak of absolute time, simply because absolute
time cannot be observed; that only clock readings, be it in the moving
reference system or the system at rest, are relevant to the determination of
time'.`Possibly I did use this kind of reasoning,' Einstein admitted, `but
it is nonsense all the same. Perhaps I couId put it more diplomatically by
saying that it may be heuristically useful to keep in mind what one has
actually observed. But on principle, \textit{it is quite wrong to try
founding a theory on observable magnitudes alone}. In reality the very
opposite happens, \textit{it is the theory which decides what we can observe}%
. You must appreciate that observation is a very complicated process. The
phenomenon under observation produces certain events in our measuring
apparatus. As a result, further processes take place in the apparatus, which
eventualIy and by complicated paths produce sense impressions and help us to
fix the effects in our consciousness. Along this whole path -from the
phenomenon to its fixation in our consciousness- we must be able to tell how
nature functions, must know the natural laws at least in practical terms,
before we can claim to have observed anything at all. Only theory, that is,
knowledge of natural laws, enables us to deduce the underlying phenomena
from our sense impressions. When we claim that we can observe something new,
we ought really to be saying that, although we are about to formulate new
natural laws that do not agree with the old ones, we nevertheIess assume
that the existing laws -covering the whole path from the phenomenon to our
consciousness- function in such a way that we can rely upon them and hence
speak of observations' '' (my emphasis).

The conversation continued for a while and at the end Einstein warned: ``%
\textit{You are moving on very thin ice. For you are suddenly speaking of
what we know about nature and no longer about what nature really does. In
science we ought to be concerned solely with what nature does}.'' Einstein
arguments are a clear support to a realistic epistomology, and I fully agree
with his views.

\subsection{Sketch of a realistic interpretation of quantum mechanics}

In this paper it is supported the hypothesis that quantum theory is a
peculiar stochastic theory. The stochasticity derives from the existence of
random fields in the vacuum. That is I assume that the vacuum is not empty
but full of fluctuating fields, Planck's constant, $h,$ fixing the scale of
the fields. More specifically every vacuum field in free space may be
expanded in plane waves, whose amplitudes are a set of statistically
independent random variables with zero mean. The square mean is such that
the average energy of one of the plane waves is $\frac{1}{2}h\nu ,$ $\nu $
being the frequency. The stochasticity of quantum theory is peculiar because
the field components of low frequency are weak but those of high frequency
are strong. This contrasts with the best known stochastic theory, namely
Brownian motion, where all components have equal strength (the spectrum
corresponds to white noise). The vacuum fields act on particles, thus
producing a random motion that departs from the classical motion. Also the
presence of matter (particles) modifies the vacuum fields, as shown for
example in the Casimir effect commented in the following.

The existence of\textit{\ virtual }fluctuating fields in the vacuum is
recognized in the most advanced form of quantum theory, namely quantum field
theory, but virtual is not a well defined concept in the first place. Thus I
assume that the fields are \emph{real}. A support to the reality of the
vacuum fields is provided by the Casimir effect\cite{Casimir}, \cite{Milonni}%
, that is an attraction between two parallel metallic plates placed in
vacuum, say at a distance $d$. In fact, due to the boundary conditions of
the electromagnetic field on a metallic surface, the (mean) energy density
and pressure of the vacuum field near the surface is different from those
quantities in free space, which gives rise to a net force per unit area, $%
F/A $, between the plates. For perfectly conducting metal and large enough
plates (i. e. $A>>d^{2}$) so that border effects are negligible, $F/A$ may
depend only on the distance, $d$, the Planck constant, $h$, and the velocity
of light, $c$, whence dimensional considerations lead to 
\[
\frac{F}{A}=K\frac{hc}{d^{4}}. 
\]
A detailed calculation (which may use purely classical electrodynamics\cite
{dice}) gives $K=$ $-\pi /480,$ the negative value meaning attractive force.
The measurement results agree with the prediction\cite{Lamoreaux}.

Elementary quantum mechanics (QM) is an approximation to field theory where
the vacuum fields do not appear explicitly. Thus QM looks like a stochastic
theory where the source of randomness is hidden. This is one of the main
obstacles for a realistic understanding of the theory. In fact in sections 2
to 4 several examples will be presented showing that quantum field theory
allows a better intuitive understanting than QM. The examples deal with the
quantum theory of the electromagnetic field (i. e. quantum electrodynamics,
QED) because that field is the most relevant in low energy phenomena,
related to atoms, molecules or condensed matter.

\section{Specific features of quantum physics}

In this paper it is proposed that the difficulties for a realistic
interpretation of quantum phenomena do not derive from the empirical facts,
or not only. Thus in the following I shall briefly revisit the most relevant
of those facts in order to analyze whether the nude empirical facts put
actual difficulties for a physical model of the microworld, independently of
the quantum formalism. The difficulties for the interpretation of the
formalism will be treated in another section.

\subsection{Discrete energy states}

As is well known the assumption that material systems may possess only
energies belonging to a discrete set was the first quantum hypothesis,
introduced by Planck in 1900. It was reinforced by the Einstein 1905
proposal that light consists of discrete pieces of energy\ (photons). In
1913 Bohr incorporated this idea to his atomic model postulating that atoms
can only exist in states having energies within a discrete set, $%
E_{0},E_{1},E_{2}$,.... The model also assumed that the absorption and
emission of light takes place with transitions between these states, the
frequency, $\nu _{jk},$ of the light related to the difference of atomic
energies by 
\begin{equation}
h\nu _{jk}=E_{j}-E_{k}.  \label{Bohr}
\end{equation}
In practice the frequencies are observed whilst the existence of energy
states is derived from eq.$\left( \ref{Bohr}\right) $. The success of Bohr's
model gave support to the hypothesis of discrete atomic energy states and
the assumption was confirmed by the experiment of Frank and Hertz in 1914.
It consisted of the scattering of electrons on mercury atoms in vapour state
with the result that, for high enough electron energies, inelastic
scattering was observed with a decrease of the electron energy by 4.9 eV.
This quantity precisely corresponds to a frequency of the mercury spectrum
via the relation eq.$\left( \ref{Bohr}\right) .$ Thus the said quantity was
interpreted as the energy difference between the ground state and the first
excited state of the atom. As a consequence of these facts, and others, it
has been fully accepted the hypothesis that the set of energy states of
atoms is discrete. Furthermore the discontinuities have been incorporated to
the quantum formalism, assuming that physical quantities should correspond
to operators (in a Hilbert space), many of them having a discrete spectrum.
Also the discreteness has been adscribed to other dynamical quantities.

The quantum discontinuities give rise to difficulties for an intuitive
understanding of quantum physics. In fact it is difficult to picture how
material systems may make transitions between two different energy states
never possessing any intermediate energy. Hints for the solution of the
difficulties is provided in the following. Firstly it is necessary to
distinguish the discreteness of the energies in the electromagnetic
radiation, that is the assumption that ligth consists of particles
(photons), from the discreteness of the energies of many-body systems like
atoms, molecules or nuclei. The nature of photons will not be discusses
here, the discontinuity of the atomic energies will be considered in the
following.

Quantum electrodynamics predicts that spectral lines are not sharp, but
possess some width. Thus Bohr's eq.$\left( \ref{Bohr}\right) $ should be
taken as an approximation, the possible atomic energies actually consisting
of a continuous set. Thus eq.$\left( \ref{Bohr}\right) $ simply recalls that
the probabilities of the energy states are strongly concentrated near some
discrete values. This solves one paradox which appears when the emission or
absorption of light is presented at an elementary level, namely the
contradiction between assuming that atomic transitions are instantaneous and
assuming that the emitted light has a sharp frequency. The fact is that the
transition has a finite duration, $\Delta t,$ and the emitted light has a
finite linewidth, $\Delta \omega ,$ fulfilling the inequality 
\begin{equation}
\Delta \omega \Delta t\gtrsim 2\pi ,  \label{1a}
\end{equation}
which is well known from classical optics. The inequality holds true for any
periodic motion and the quantum formalism also predicts it. Indeed sharp
energies of atoms appear only in a calculation to lowest order of
approximation (i. e. in the limit when the electron charge $e\rightarrow 0)$%
. However when radiative corrections of quantum electrodynamics are taken
into account the calculation leads to spectral lines with a finite width.
The \emph{corrections} are small and \emph{may be neglected in elementary
calculations} but \emph{they are essential for a realistic interpretation. }%
This is a typical example of how the emphasis on the simplicity of the
calculations, rather than the clarity of the concepts, has the consequence
that quantum mechanics appears as counterintuitive. A realistic picture of
the atomic emission is possible assuming that light is emitted in a
continuous process lasting a time $\Delta t$ that fulfils the inequality $%
\left( \ref{1a}\right) ,$ the total energy of atom plus light being
conserved at all instants of time. Indeed this fits with the quantum
electrodynamical evolution equation of the atom coupled to the
electromagnetic field. Of course the realistic picture is incomplete because
some explanation should be found for the relation between frequency and
energy, which is quite strange from a classical perspective. But the lack of
a realistic model (till now) does not imply that a model is impossible.

Similar arguments may be used in order to understand the quantization of
angular momentum, as shown for instance in the Stern-Gerlach experiment. In
popular expositions the experimental results are presented as if all atoms
arrive at one amongst two sharp lines in a screen. Then it is difficult to
reach a picture of what is taking place in the interaction between the atom
and the inhomogenous magnetic field. However the truth is that what appears
in the screen are two wide spots, something much less counterintuitive. It
is the case that an accurate quantum mechanical treatment of the experiment
precisely predicts that\cite{SternGerlach}. A picture of the phenomenon may
be reached assuming that during the interaction of the atom with the
magnetic field some fluctuation and dissipation takes place which tends to
alineate (approximately) the atomic magnetic moment with the field.

\subsection{Heisenberg uncertainty principle}

The Heisenberg uncertainty principle is the most frequently quoted evidence
of the dramatic splitting between classical and quantum physics. In fact the
principle appears in popular writings like a kind of mysterious property of
our world. I shall not discuss here the general principle dealing with
conjugate dynamical variables. I will restrict attention to the
experimentally proved impossibility of determining simultaneously the
position and the velocity (or the momentum) of a particle. This implies that
it is not possible to prepare a particle with both position and velocity
sharply defined, and also that no measurement may provide the values of both
these quantities at the same time. A consequence is the impossibility of
finding empirically the path of a particle. In the following it is shown
that a realistic interpretation is possible by analogy with what happens in
Brownian motion. A Brownian particle possesses a highly irregular path whose
instantaneous velocity cannot be measured (with ordinary, macroscopic
set-ups). Only the mean velocity, $\stackrel{-}{\mathbf{v}},$ during some
time interval may be measured, that is 
\[
\stackrel{-}{v}=\frac{\left| \Delta \mathbf{r}\right| }{\Delta t}, 
\]
where $\left| \Delta \mathbf{r}\right| $ is the distance between the initial
and final positions in the time interval $\Delta t.$ On the other hand there
is a relation, derived by Einstein in 1905, between the expected value of
the square of the distance, $\left| \Delta \mathbf{r}\right| ^{2}$, and the
time interval, $\Delta t$. That is 
\[
\left\langle \left| \Delta \mathbf{r}\right| ^{2}\right\rangle =D\Delta t, 
\]
where $D$ is called diffusion constant and $\left\langle {}\right\rangle $
means ensemble average, that is the average over many measurements involving
the same time interval. If we eliminate $\Delta t$ amongst the two
equalities we get 
\[
\left\langle \left| \Delta \mathbf{r}\right| ^{2}\right\rangle =\left\langle 
\stackrel{-}{v}^{2}\right\rangle \Delta t^{2}\Rightarrow \left\langle \left|
\Delta \mathbf{r}\right| ^{2}\right\rangle \left\langle \stackrel{-}{v}%
^{2}\right\rangle \simeq D^{2}, 
\]
a relation having some similarity with the Heisenberg inequality. In
conclusion a plausible interpretation of the Heisenberg principle is that
the quantum motion possesses a random component having similarity ( not
identity!) with Brownian motion. This similarity has been the basis for the
development of stochastic mechanics\cite{stochastic}, which provides an
intuitive picture of some typically quantum phenomena. However this theory
present difficulties that will not be discussed here.

The Heisenberg principle becomes an obstacle for a realistic interpretation
of quantum mechanics when the empirically found practical difficulty (or
impossibility) of simultaneous knowledge of position and velocity is
elevated to the category of an ontological statement: ``\emph{Trajectories
of quantum particles do not exist''.}

\subsection{The apparent lack of objective properties.}

In classical physics it is an implicit hypothesis that any observation or
measurement just reveals (`removes a veil') a property which exists
objectively with independence of any observation. In quantum mechanics this
seems to be untrue. Let us clarify the motivation for that belief with an
example. We consider a physical system possessing three observable
properties which I shall label $A$, $B$ and $C.$

I will assume that the observables $A$ and $C$ may be measured in the same
experiment and similarly for $B$ and $C$, but for some reasons $A$ and $B$
cannot be measured with the same experimental arrangement. Then with
repeated measurements in identically prepared systems it is possible to
obtain the joint probability distribution for the results of the former
measurement, which I will represent by the density, $\rho \left( a,c\right)
, $ that the observable $A$ takes the value $a$, and the observable $C$ the
value $c$. Similarly we may obtain $\rho \left( b,c\right) ,$ but it is not
possible to obtain empirically a joint probability density $\rho \left(
a,b\right) $ because we cannot measure $A$ and $B$ simultaneously$.$ Up to
here no problem arises, everything agrees with the intuition.

Now if we think that the measurement just reveals preexisting values of the
observable quantities we are compelled to assume that, in every state, the
system possesses the values $a,b$ and $c$, independently of any observation
or measurement. More generally the preparation procedure will lead to a
state with a joint probability distribution, $\rho \left( a,b,c\right) ,$
for the three observables. If this is the case the joint probabilities for
two observables should be the marginals of the former distribution, that is 
\begin{equation}
\rho \left( a,c\right) =\int \rho \left( a,b,c\right) db,\;\rho \left(
b,c\right) =\int \rho \left( a,b,c\right) da.  \label{marginals}
\end{equation}
However it has been shown in some experiments that there are particular
cases of states and observables where no (positive) joint probability
density $\rho \left( a,b,c\right) $ exists such that the marginals eqs.$%
\left( \ref{marginals}\right) $ agree with the empirical results. The
non-existence of a joint probability fulfilling eqs.$\left( \ref{marginals}%
\right) $ in general is predicted from the quantum formalism, and it is the
essential content of the Kochen-Specker theorem (see e.g. Mermin\cite{Mermin}%
.) As in the case of the Heisenberg principle, the practical impossibility
has been raised to the rank of an ontological statement: \emph{``Physical
systems do not possess properties independently of measurements.''}

Is that statement justified?. It is not\cite{Bell65}. What the experiments
have shown is that the observed properties depend not only on the state of
the system but on the whole experimental set-up. In fact, we may assume that
physical systems possess some properties, called `elements of reality' by EPR%
\cite{EPR}, which in specific experimental set-ups give rise to observable
quantities, but the obsevables may not exist independently of the
experiment. Actually a similar situation also happens in classical physics,
as for instance when we play dice. If we get a number, say 2, we cannot
claim that the value 2 was preexistent to our experiment. In fact the result
2 is actually `created' by the experiment of throwing the dice. Returning to
quantum physics, there is a simple explanation for the frequent inexistence
of properties independent of measurements (some particular properties do
exist, for instance the rest mass of particles). We may assume that the
measured properties are \emph{contextual}, that is they depend not only on
the state of the system but on the whole experimental context. This point
was correctly emphasized by Bohr and, in my opinion, solves all problems of
interpretation which might follow from the Kochen-Specker theorem. (Of
course the theorem provides some quantitative statements which should be
explained, but here I am addressing the question whether the practical
impossibility of getting joint probabilities prevents a realistic
interpretation.)

The real difficulty arises when people attempts to reach conclusions which
go beyond what follows from the facts. Indeed we can state that \emph{some
properties} do not exist independently of measurements \emph{in some
particular instances}, but we should not extrapolate telling that in nature 
\emph{there are no properties} independent of the observation. This absurd
extrapolation was correctly criticized by Einstein with his celebrated
rhetorical question\emph{\ ``Is the moon there when nobody looks?''}\cite
{Einstein}\emph{.}

One might ask why in the microscopic domain it is frequent that values of
the observables are created by the experiments whilst this situation is rare
at the macroscopic level. An explanation may be as follows. In the
macroscopic world we may study systems with instruments more fine than the
object to be studied. E. g. we may look at the interior of an orange using a
knife. In the microscopic domain any macroscopic equipment used for the
study of atoms will consist of atoms. This makes our knowledge less direct
in the micro than in the macroscopic domain, and more dependent on the
context.

The fact that the measurement cannot be understood as simply revealing the
values of preexising properties has led to the introduction of some
`postulates of the measurement' in the standard quantum formalism as
discussed in the next section.

\subsection{Statistical character}

Typical experiments are affected by statistical errors even in the classical
domain. That is the same experiment performed in similar conditions may give
rise to (slightly) different results. For this reason it is a standard
practice to report the results of measurements accompanied by an uncertainty
interval. In the macroscopic domain the uncertainty is attributed to the
difficulty in controlling a very large number of parameters, with the
consequence that never (or rarely) an experiment may be repeated in exactly
the same conditions. In any case it is usual that the uncertainty is only a
very small fraction of the measured quantity. In contrast in the microscopic
domain it is frequent that the uncertainties are of the same order than the
measured result. This is equivalent to say that the same experiment may give
rise to a number of different results, every one with some probability. For
this reason it is said that any theory of the microworld should be
statistical, that is giving predictions in the form of probabilities of
several different outcomes. However, at a difference with macroscopic
(classical) physics, in quantum physics the probabilities are usually not
atributted to lack of control in the experiment.

The current wisdom is that quantum probabilities are radically different
from the classical, ordinary life, probabilities. The latter derive from
incomplete knowledge (`ignorance'), maybe unavoidable, about the truth of
some assertion. For instance we may attach a probability 1/2 to the
appearance of head in throwing a coin, because we cannot control all
relevant variables in the experiment. In contrast it is assumed that quantum
probabilities are quite different, that they derive from a lack of strict
causality of the natural laws, that is the fact that different effects may
follow to the same cause. This is usually called the \emph{fundamental or
essential probabilistic} character of the physical laws. Again a practical
difficulty has been raised to the rank of an ontological statement: \emph{%
``Natural laws are not strictly causal''.}

Einstein disliked that assumption and strongly criticized it, as shown by
his celebrated sentence ``\emph{God does not play dice}''. I understand very
well Einstein\'{}s opinion. For him the rational understanding of nature was
a kind of religion. As more loose (strict) are the natural laws smaller
(greater) could be our rational understanding of nature. Accepting a weak
causality is like accepting poor science. Nevertheless there are people
happy with the absence of determinism implied by the nonexistence of strict
causality. For instance some claims have been made that the quantum lack of
determinism may explain human freedom. This question lies outside the scope
of this paper and will not be further commented.

But I do not support determinism in the mechanicistic view of Laplace. In my
opinion quantum mechanics is a stochastic theory. There are strictly causal
laws, but there is also an universal noise (the random vacuum fields) which
permeates everything and prevents any practical determinism in the evolution%
\textrm{\ }(see the Introduction section). Strict causality combined with
stochasticity (randomness) is in practice indistinguashable from essential
probability, and the former is more plausible. In order to clarify the
matter let us think again in Brownian motion. Under macroscopic observations
the random motion of a Brownian particle may appear as lacking causality,
but we assume that, taking into account the molecules of the liquid where
the particle is immersed, the whole motion is governed by Newtonian
dynamics, which is causal.

\subsection{Wave-particle duality}

The assumption that all quantum entities have a dual nature, particle and
wave, is the source of most difficulties for an intuitive understanding of
quantum mechanics. But if we do not want to destroy the basic properties of
space, the wave-particle duality really involves a contradiction. In fact\ 
\emph{particle} means something localized, \emph{wave }means something
extended. More precisely, particle (wave) means smaller (much greater) than
some reference length, usually a few times the size of an atom. For this
reason it is bizarre to say that an atom (with radius about 10 nm) passes
simultaneously through two slits (distant about 1$\mu $m). Thus it is not
strange that for some people, like Feynman, the interference experiments
contain all the mysteries of quantum mechanics. The problem posed by the
wave-particle duality for a realistic interpretation of the quantum
phenomena is certainly big. It will be considered elsewhere, but in the
following I sketch a possible solution. We may assume that in nature there
are both particles and fields (waves), the particle behaviour of fields
deriving from the interaction with particles and the wave behaviour of
particles from the interaction with fields. The difference with the
macroscopic world, where there are also particles and fields, is that
interactions are more relevant and complicated in the microscopic domain.

I think that the electrons (or protons, neutrons, atoms, molecules) are
particles, whilst radiation consists of waves. `Photons' are not particles
but mathematical constructs useful for the description of some phenomena\cite
{Lamb}. Then, how may we interpret the interference experiments where we
observe fringes typical of waves, but these fringes appear as sets of
localized events which are typical of particles? In the case of radiation
the interference may be easily understood in classical terms, the problem is
the particle behaviour in the detection. The opposite is true for particles
like atoms. Its localized detection is easy to understand but their
interference puts the problem. Let us study the two cases separately.

The detection of `individual photons' in a photographic plate is due to the
atomic nature of the plate. In this case saying that radiation are particles
because they give rise to individual blackened grains is like saying that
wind is corpuscular because the number of trees falling in the forest is an
integer. Of course in both cases, the photo and the forest, there is a
random element. It is obvious for the wind but there is also a random
element in the radiation: the quantum noise or quantum vacuum fluctuations.
The detection process in a photon counter may be explained as a transfer of
energy from the field to individual atoms or molecules, this producing
excitation that in some cases results in one count. I believe that the
vacuum fields play a relevant role in this process.

The wave behaviour of neutrons, atoms or molecules, for instance in the
two-slits experiments, is more difficult to understand. We might assume that
it is caused by the vacuum fields, mainly the electromagnetic radiation. For
instance let us consider a metallic plate with two holes. The
electromagnetic vacuum fields near the plate will be different from the
fields when the plate is not present (the difference gives rise to the
Casimir effect, see the Introduction section above). Thus it is plausible to
assume that the waves traveling from the left (right) of the plate give an
interference pattern at the right (left) side. Thus any particle with net
charge (e. g. an electron) or charged parts (a neutron or an atom) crossing
one of the holes will have a motion modified by the action of those waves,
which provides a \emph{qualitative} understanding of the particle
interference. The picture has some similarity with the old proposal by L. de
Broglie (the pilot wave theory) or to the picture offered by `Bohmian
mechanics'\cite{Bohmian}, but there are important differences. Firstly in
our view there is a clear physical entity causing the interference, namely
the vacuum fluctuations, whilst the particle remains localized all the time.
Secondly there is a random element which is not present in Bohmian
mechanics. Of course any model resting upon these ideas would be rather
involved and it is not easy to understand why quantum mechanics provides so
simple rules for the \emph{quantitative} prediction of the empirical
results. In fact the interference experiments with particles put a big
challenge for a realistic interpretation of quantum theory.

\subsection{Conclusions}

The analysis of the most characteristics quantum phenomena leads me to
emphasize a point that is crucial for the attempt of reaching a picture of
the quantum world. \emph{The difficulties for a realistic interpretation of
quantum mechanics may derive from a number of unneeded assumptions, adhered
to the minimal quantum formalism for historical reasons. In some cases the
difficulties are caused by an excess of idealization in the interpretation
of the experiments.} An example of the former is the claim that quantum
particles have no trajectories. An instance of the latter is the usage of
idealizations and the adscription of any deviation from the idealized result
to accidental errors. It is assumed that this contributes to the clarity but
in my opinion it is the opposite, it contributes to misunderstanding. It is
true that the method may simplify the teaching of \emph{how to use} quantum
mechanics, but it puts a strong obstacle for getting an intuitive picture of
the quantum world. A typical example, commented above, is the use of first
order perturbation theory in the study of emission or absorption of light,
which hides the fact that the formalism (quantum electrodynamics) predicts a
continuous evolution of atom plus field.

\section{Critical comments on current interpretations}

As said above, Heisenberg quantum mechanics was proposed as an abstract
formalism without any physical picture behind. Bohr justified the absence of
a model and, on this basis he elaborated the `Copenhagen interpretation'\cite
{Bohr}. Later on several modifications or novel interpretations have appeared%
\cite{Jammer}, \cite{Laloe}. Relevant papers up to 1983 are reprinted in in
a book by Wheeler and Zurek\cite{WZurek}. I shall comment briefly on the
most popular interpretations in the following.

For the sake of clarity I will illustrate the comments with the celebrated
`Schr\"{o}dinger cat' \emph{gedanken} experiment\cite{Schrödinger}. It
consists of a box containing a radiactive atom and a cat toghether with a
device that kills the cat, say instantaneously, when the atom decays. I will
assume that both the atom in the excited state and the live cat are put
inside the box at time $t_{1}.$ The question is what may be said about the
atom and the cat at times $t>t_{1}$. In particular, what is the prediction
of quantum mechanics for the states of both the cat and the atom when the
box is open at time $t_{2}$. Any person with knowledge of the law of
radiactive decay, but ignorant of quantum mechanics, would claim that the
probability of being both the cat alive and the atom excited at time $t\in
\left[ t_{1},t_{2}\right] $ is 
\begin{equation}
P(t)=exp\left[ -\lambda \left( t-t_{1}\right) \right] ,  \label{1}
\end{equation}
$\lambda ^{-1}$ being the mean lifetime of the atom. In particular the
probability at the moment of opening the box will be $P(t_{2})$, eq.$\left( 
\ref{1}\right) $. This may be named the response of a `naive realist'. In
contrast, the answer of an educated quantum physicist will depend on the
interpretation that she/he supports, as I comment in the following.

\subsection{Copenhagen interpretation}

According to the Copenhagen interpretation (CI)\ the referent of quantum
mechanics is not the material world but the experiments. That is the theory
deals with the relations between the world and the observers. As Bohr put it
``the finite magnitude of the quantum of action prevents altoghether a sharp
distinction being made between a phenomenon and the agency by which it is
observed''\cite{Bohr}. Thus CI is close to the pragmatic approach as
commented in the Introduction section. According to this approach we should
not make assertions about the bodies, but about the results of possible
observations or measurements. Thus a sentence like $``$the probability that
the atom \emph{is} in the excited state at time $t"$ is considered
meaningless. A meaningful assertion should be something like ``if we perform
a measurement of the state of the atom at time $t,$ the probability that we
get the result `excited' is given by eq.$\left( \ref{1}\right) ".$ The
approach is instrumentalist and it might be called a `protocol for the use
of the quantum formalism' rather than an interpretation. Bohr elaborated a
philosophical background with the introduction of the `complementarity
principle' and the `correspondence principle', in order to solve two
theoretical difficulties of the formalism. Firstly it is unsatisfactory that
quantum mechanics applies only to the microscopic world whilst the
macroscopic one is governed by classical theories. Bohr\'{}s solution to
this problem was to assume that there is a smooth transition, quantum laws
approaching the classical ones in the limit when Planck constant becomes
negligible, formally when $
\rlap{\protect\rule[1.1ex]{.325em}{.1ex}}h%
\rightarrow 0.$ This is the essential content of the \emph{correspondence
principle, }that Bohr also applied to several instances deriving some
relevant results. The second theoretical problem was the existence of
apparent contradictions, in particular those derived from the fact that
quantum entities behave sometimes like particles and other times like waves.
In order to solve that problem Bohr proposed the \emph{complementarity
principle}, which stresses the incompatibility between causal laws and
spacetime description, due to the finite (nonzero) value of the quantum of
action. After that he showed that there is no contradiction in practice
because the behaviour of the quantum entities does not derive from the
microscipic system alone, but also depends on the full context\cite{Bohr},
including macroscopic measuring devices.

The rules of CI for the use of quantum mechanics are to some extent
independent on the two mentioned Bohr's principles, and I will comment only
on the rules. CI assumes (or at least it does not reject the assumption)
that macroscopic bodies have objective properties (that is independent of
any measurement) and its evolution is governed by the laws of classical
physics. Thus it is meanigful to ask whether a cat is either alive or dead
at any time. A more difficult question is whether we are allowed to assign a
probability to every one of these possibilities. The application of quantum
mechanics, with the CI rules, to the ``cat experiment'' is that for $t\in
\left( t_{1},t_{2}\right) $ the atomic state should be represented by the
state vector 
\begin{eqnarray}
&\mid &\psi \left( t\right) \rangle =c_{g}(t)\mid g\rangle +c_{e}(t)\mid
e\rangle ,  \label{2} \\
c_{e}(t) &=&\sqrt{exp(-\lambda \left( t-t_{0}\right) )},c_{g}(t)=\sqrt{%
1-exp(-\lambda \left( t-t_{0}\right) )},  \nonumber
\end{eqnarray}
where $\mid g\rangle $ ($\mid e\rangle )$ is the state vector of the atom in
the ground (excited) state. Now there are two possibilities depending on
what is supposed to be a measurement: 1) If we assume that the actual
measurement takes place when the box is open, then quantum mechanics says
nothing about the atom and the cat for times $t\in \left( t_{1},t_{2}\right)
.$ At time $t_{2}$ it predicts that the probability of both the cat being
alive and the atom being excited is given by the modulus square of the
amplitude $c_{g}(t),$ eq.$\left( \ref{2}\right) ,$ which precisely agrees
with the naive prediction $P(t_{2})$, eq.$\left( \ref{1}\right) .$ 2) We
might assume that the cat, being a macroscopic system, may act as measuring
device. In this case, CI tells us that, for any time $t\in \left(
t_{1},t_{2}\right) ,$ the probability of both the cat being alive and the
atom excited is eq.$\left( \ref{1}\right) $. The latter interpretation (the
cat as measuring device) is consistent with the fact that, if the cat is
found dead at time $t_{2},$ a careful study of the corpse (involving
macroscopic manipulations) might determine the time of death, say $t_{d}$.
This would allow reconstructing the whole history: The cat was alive and the
atom excited until $t_{d}$. We must assume that, if a similar experiment is
performed many times, the distribution of times $t_{d}$ would converge to an
agreement with the probability eq.$\left( \ref{1}\right) .$

It is interesting the Bohr approach to the problem of the `state vector (or
wave function) collapse'. This is the discontinuous change of the state
vector when a measurement is made, e. g. a change from eq.$\left( \ref{2}%
\right) $ to $\mid \psi \rangle =\mid g\rangle ,$ at the time of opening the
box. In our example we may naively believe that the collapse is just a
change of our information as a result of the observation. However Bohr
strongly opposed to the belief that the \emph{wave function just represents
our information about the system, }with the implicit consequence that this
information may be incomplete.\emph{\ }See section 4.1 for a more detailed
discussion of the completeness question.

\subsection{John von Neumann}

CI is very good from the practical point of view and avoids any bizarre
assumption (which is not the case in more elaborated interpretations
commented below.) The problem with the CI is that it creates what has been
called an `infamous boundary'\cite{Wick}, that is a discontinuity between
micro and macrophysics. The former should be studied within quantum
mechanics, the latter using classical physics. In order to remove the
boundary and get an interpretation where quantum mechanics is valid also for
macroscopic systems, John von Neumann\cite{von Neumann} introduced a theory
of measurement and even he gave a model for it. His approach has been also
currently named Copenhagen interpretation, but this is misleading because
von Neumann\'{}s interpretation is different from Bohr's as a matter of
principle. However it was an elaboration of the Copenhagen interpretation
rather than an alternative, which may justify the name. For short I shall
label it MCI with M standing for modified or measurement. MCI has been
supported in most papers and books of quantum mechanics until around 1980.

The modification introduced by von Neumann with respect to Bohr was to take
seriously the assumption that quantum mechanics is the universal theory and
classical theories are just approximations. Thus he proposed studying the
measurement within quantum mechanics and made a model involving the coupling
of the microscopic system with the measuring apparatus. However this gave
rise to a number of difficulties that will be commented below, but
previously we clarify the matter studying the application of von Neumann's
ideas to the cat experiment.

In MCI both the cat and the atom should be treated as quantum objects.
Therefore eq.$\left( \ref{2}\right) $ is no longer appropriate and we should
represent the state of the whole system, atom plus cat, by 
\begin{equation}
\mid \psi \left( t\right) \rangle =c_{g}(t)\mid g\rangle \mid deadcat\rangle
+c_{e}(t)\mid e\rangle \mid livecat\rangle .  \label{3}
\end{equation}
Of course one may point out that a dead cat does not correspond to a pure
state to be represented by the vector $\mid deadcat\rangle .$ Indeed there
may be very many quantum states corresponding to a dead cat and similarly
for a live cat. However this is not a real problem because MCI assumes that
any physical system is associated to a well defined state vector. (When the
appropriate state vector is not known we should use a probability
distribution over those vectors, which may be formalized via a density
matrix. But for simplicity we may use a single state vector as in eq.$\left( 
\ref{3}\right) $)

Eq.$\left( \ref{3}\right) $ represents a typical `entangled state', a name
introduced by Schr\"{o}dinger in 1935\cite{Schrödinger}. If CI had been
modified with the assumption that state vectors actually represent
statistical ensembles this would have lead to the ensemble interpretation,
to be commented below. However the mainstream of the scientific community
rejected it and supported the `completeness' of quantum mechanics, in the
sense that \emph{the state vector represents the actual state of an
individual physical system, as opposed to a statistical ensemble}. With this
assumption the MCI leads to bizarre consequences, which was the point that
Schr\"{o}dinger\cite{Schrödinger} attempted to stress with his cat example.
Indeed for many people it is impossible to understand the meaning of a state
represented by a superposition of alive and dead cat. Is it something
intermediate between life and death?.

The problem is not only the highly counterintuitive character of the
superpositions of macroscopic systems, it is the disagreement with empirical
evidence. Indeed it is the case that those macroscopic superpositions cannot
be manufactured in practice (there is a lot of literature about the actual
preparation of `Schr\"{o}dinger cats', but they always involve mesoscopic
rather than truly macroscopic systems.) Thus it seems that the quantum
evolution (the Schr\"{o}dinger equation) is violated at the macroscopic
level. This has been called the problem of the \emph{objectification or
individuation}\cite{Mittelstaedt}. That is the fact that \emph{a particular
value is obtained in the measurement amongst several possible values,}
something not predicted by the quantum formalism except if an explicit
postulate is included. This postulate forces us to change the state vector
at the time of measurement, a change usually called the `state vector, or
wave function, collapse'. That change is not predicted by the
Schr\"{o}dinger equation and in fact it precludes the validity of that
equation during measurements. In the CI the collapse was just a change of
the mathematical representation needed for the analysis of the experiment.
However in MCI it becomes a real physical change because it is assumed that
the state vector corresponds to an individual system (rather than a
statistical ensemble). In the next subsection I discuss possible solutions
that have been proposed.

A problem related to the objectification is the existence of quantum jumps%
\cite{Belljumps}, the typical example being the decay of a radioactive atom.
For instance an atom of uranium 238 may remain as such during million years
but, at some unexpected time, it decays with the emission of an alpha
particle (a nucleus of helium 4). The sudden decay (within a small fraction
of one second) apparently contradicts Schr\"{o}dinger equation. People like
to say that the observation of the (spontaneous) decay is a particular case
of measurement and therefore the problem of the quantum jumps becomes an
example of objectification after a measurement. In my opinion however there
is a clear difference between objectification and quantum jump. The former
is more properly the `disentanglement' of an entangled state involving
macroscopic bodies. For instance the fact that we see the cat alive or dead
at the time of opening the box in the Schr\"{o}dinger cat example. In
contrast a quantum jump refers to a discontinuous change of a microscopic
system. In any case the difficulties with both objectification and jump may
be solved simultaneously, for instance in either hidden variables or
collapse theories, the latter discussed in the next subsection and the
former later on.

\subsection{The objectification and the quantum jumps problems}

There are a variety of proposed solutions to the objectification problem. In
the original (Bohr) Copenhagen interpretation there is no real problem: the
Schr\"{o}dinger equation is just a mathematical tool able to relate the 
\emph{preparation} of a (microscopic) system to \emph{measurements} made on
it. The wave function (or state vector) is just a convenient form of dealing
with the probabilities involved. That is a preparation gives rise, after
some time, to a set of probabilities when the system is placed in an
appropriate experimental context. The objectification is a change due to the
measurement. But the change must be \emph{postulated} because the
interaction between the microscopic system and the macroscopic apparatus can
be described neither by quantum nor by classical theories in CI. The
objectification problem also does not exist in the `many worlds
interpretation'(MWI) that will be discussed in the next subsection.

From the time of von Neumann\'{}s book\cite{von Neumann} (1932) until around
1980\'{}s, and for a fraction of the scientific community until today, the
MCI has been the most popular interpretation. For this reason a very large
number of papers and books have been devoted to propose possible solutions
to the objectification problem.

John von Neumann pointed out that the measurement only finish when a (human)
observer is conscious of the result of the experiment. This would solve the
objectification problem if we assume that the mind is not governed by
quantum mechanics.The proposition was also supported by London and Bauer\cite
{London} and commented by Wigner\cite{Wigner}. In the cat experiment, this
seems to imply that the cat really dies when we look at the box after
opening it, or even when we are informed by another person of the result of
the experiment (this leads to the `Wigner's friend' paradox.) The solution
dislikes many people.

In practice many authors (maybe not too fond of the subtleties of
foundational questions) accepted a kind of peaceful coexistence of the two
(contradictory) postulates: the Schr\"{o}dinger equation and the quantum
theory of measurement. This attitude however has been strongly critiziced by
philosophers of science\textrm{\ }like Karl Popper\cite{Popper} or Mario
Bunge\cite{Bunge}. In particular the latter stresses that a physical theory
should not include a general theory of measurement, but particular recipes
or protocols for every specific measurement. This is most clear in
chemistry. There are recipes for, say, the preparation of pure alcohol or
the analysis of water of a river. However it would be absurd to search for a
`general theory of preparation or analysis' in chemistry. In my opinion the
same is true in physics, including quantum physics. Actually the existence
of a `theory of measurement' is peculiar, it does not exist in any other
theory in physics (or more generally in natural science). It is true that
from a philosophical (epistemological) point of view any theory requires
some assumptions for the connection with the results of observations or
experiments. For instance in classical mechanics we use the concepts of
time, space, particle, isolated system, etc., and there are rules telling us
how these concepts should be related to the (mathematical) formalism.
However it would be absurd to search for a `general theory of preparation or
measurement' in physics, including quantum physics.

A proposal that has been popular since around 1985 is to modify the
Schr\"{o}dinger equation in such a way that the change fulfil two
consistency requirements: 1) For microscopic systems it produces an
extremely weak, practically undetectable, modification in the evolution of
the wave function, and 2) For macroscopic systems it gives rise to a rapid
disentanglement, that is an evolution from any superposition to a single
term. There have been several explicit models of this type, called `collapse
theories'. The most satisfactory has been proposed by Ghirardi, Rimini, and
Weber in 1985, and usually referred to as the GRW theory\cite{GRW}, \cite
{GPearleR}. At present, it involves phenomenological parameters that, if the
theory is taken seriously, acquire the status of new constants of nature.
There have been also attempts at deriving the parameters from fundamental
arguments, like the action of gravitational forces (effects of general
relativity.)

In spite of their phenomenological character, the collapse theories have
relevance since they have made clear that there are new ways to overcome the
difficulties of the quantum formalism. Moreover, they have allowed a clear
identification of the formal features which should characterize any unified
theory of micro and macro processes. Last but not least, collapse theories
qualify themselves as rival theories of quantum mechanics and one can
identify some of their physical implications which, in principle, would
allow crucial tests discriminating between the two. This possibility, for
the moment, seems to require experiments which go beyond the present
technological possibilities. I shall not review here the collapse theory,
which would lead far from the main purpose of the paper. The interested
reader may look at a good review by Ghirardi\cite{Ghirardi}.

\subsection{Many-worlds}

The many worlds interpretation (MWI) offers a radical solution to the
objectification problem, it assumes that objectification never takes place.
That is, the evolution of an isolated system is always governed by the
Schr\"{o}dinger equation. Now no system involving a macroscopic body may be
completely isolated, so that in the study of its evolution we should
consider the wave vector of the whole universe. In particular, in the cat
experiment we should include, in addition to the atom and the cat, also the
box, the human observer and everything else. Thus eq.$\left( \ref{3}\right) $
should be replaced by 
\begin{eqnarray}
&\mid &\psi \left( t\right) \rangle =c_{g}(t)\mid g\rangle \mid
deadcat\rangle \mid world-g\rangle  \nonumber \\
+c_{e}(t) &\mid &e\rangle \mid livecat\rangle \mid world-e\rangle ,
\label{4}
\end{eqnarray}
where $\mid world-g\rangle $ represents the rest of the world associated to
the atom in the ground state and the cat dead, and similarly for $\mid
world-e\rangle .$ Eq.$\left( \ref{4}\right) $ seems to tell that there are
two copies of the human observer and of the whole world. In the latter copy
the observer sees the cat alive and the atom excited, in the former she/he
sees the cat dead and the atom in the ground state. The state vector of the
universe is a linear combination of these copies.

MWI is the unavoidable end of the logical path if we believe that quantum
mechanics (as defined by the standard postulates excluded those of
measurement) is universally valid. It was initially proposed by Everett\cite
{Everett} with the name of `relative states interpretation' and elaborated
later by de Witt\cite{de Witt}, who introduced the name `many worlds'. The
aims of MWI are: 1) retain the unrestricted validity of the quantum
formalism, 2) remove the need of the state vector collapse, 3) remove the
need of an external observer, and 4) derive the Born rule\cite{Marchildon}.
The latter is the rule for finding the probabilities of the different
possible outcomes as a result of a measurement.

Apart from the difficulty of understanding the real meaning of `multiplicity
of worlds', the main problem of MWI is to reproduce the Born rule without
introducing any explicit probabilistic postulate. The standard approach to
do that is the theory of decoherence\cite{Omnes}, \cite{Joos}. Decoherence
is the evolution predicted by quantum mechanics when a system possesses very
many degrees of freedom, as is the case for a measuring device in contact
with the environment. It involves a loss of coherence which leads from a
state vector (representing a pure quantum state) to a density matrix, as a
result of the interaction with the environment. That density matrix is
approximately diagonal in an appropriate basis (the preferred basis), so
that it looks like a probability distribution defined on a set of pure
states, as is exhibited in the example eq.$\left( \ref{3a}\right) $ below.
In the context of MWI the density matrix may be seen as coming from taking
the partial trace over those degrees of freedom which are not of interest.
For instance if we take the partial trace, with respect to the world states,
of the (idempotent) density matrix associated to the state vector eq.$\left( 
\ref{4}\right) $ we get with very good approximation 
\begin{equation}
Tr_{world}\left| \psi \rangle \langle \psi \right| \simeq \left|
c_{g}(t)\right| ^{2}\left| g-d\rangle \langle g-d\right| +\left|
c_{e}(t)\right| ^{2}\left| e-l\rangle \langle e-l\right| ,  \label{3a}
\end{equation}
where $\mid g-d\rangle $ is short for $\mid g\rangle \mid deadcat\rangle $
and $\mid e-l\rangle $ for $\mid e\rangle \mid livecat\rangle $ and the
orthogonality of the state vectors $\mid world-g\rangle $ and $\mid
world-e\rangle $ has been taken into account. Eq.$\left( \ref{3a}\right) $
is mathematically identical to the representation of the quantum state of
the atom plus the cat that we should use when we do not know its actual
state, and consequently we attribute the probability $\left| c_{g}(t)\right|
^{2}$ to the atom being in the ground state and the cat dead, and $\left|
c_{e}(t)\right| ^{2}$ the probability of the alternative. The question, to
be discussed below, is whether eq.$\left( \ref{3a}\right) $ actually
corresponds to a mixture or not.

Actually decoherence theory is more involved than it may appear from our
example. Firstly we should consider very many terms in the sum which
represents the quantum state of the world, rather than only two as in our
simplified example eq.$\left( \ref{4}\right) .$ Also there is an ambiguity
in the world state vector because, it being a linear combination of (tensor)
products of state vectors, it could be written in many different forms
depending on the choice of basis in the Hilbert space. This leads to the
problem of the preferred basis, whose solution is one of the achievements of
decoherence theory. I shall not discuss here in more detail the different
approaches and the technical issues of the MWI and decoherence, and refer to
the vast literature on the subject (see e. g.\cite{Marchildon} and
references therein). Related to decoherence is the ``consistent histories''
approach\cite{Griffiths}, which will not be commented here.

MWI has the virtue that it makes quantum mechanics a selfconsistent theory
resting upon a simple hypothesis: its universal validity. In this respect it
is superior to the old-fashioned CI and MCI. However as usually understood
it leads to a rather bizarre picture of the world. For the sake of clarity I
will consider a measurement with reference to eq.$\left( \ref{3a}\right) $,
although now `cat' means the macroscopic measuring device able to suffer an
irreversible evolution. Once MWI plus decoherence theory leads to a reduced
density matrix like eq.$\left( \ref{3a}\right) ,$ it seems plausible to
interpret it as representing a mixture, $\left| c_{g}(t)\right| ^{2}$ and $%
\left| c_{e}(t)\right| ^{2}$ being probabilities in the usual sense of
mathematical measures of information. However this interpretation is not
compatible with the assumption that quantum mechanics is complete. That is,
the hypothesis that eq.$\left( \ref{3a}\right) $ represents a mixture is not
compatible with the assumption that the state vector of the universe
corresponds to an individual world (although with many branches), rather
than a statistical ensemble of possible worlds. However it is irrelevant in
practice whether we assume that the world state vector represents complete
or incomplete information. In fact a \emph{detailed knowledge} of that state
vector would always lie beyond the human capabilities. Therefore the
assumption that eq.$\left( \ref{3a}\right) $ represents an actual mixture,
and quantum mechanics is incomplete, is in my opinion most plausible.

In contrast, the conjunction of assuming universal validity (i. e. MWI) and
completeness of quantum mechanics leads to the extravagant view that there
are many parallel worlds\cite{Marchildon}. I think that this belief is
unnecessary. Actually the view rests on what has been termed \emph{a
Platonic paradigm} by M. Tegmark\cite{Tegmark}, who defines it as follows:
``The outside view (the mathematical structure) is physically real, and the
inside view and all the human language we use to describe it is purely a
usefull approximation for describing our subjective perceptions.'' The
mathematical structure referred to by Tegmark is the formalism of quantum
mechanics. Thus the Platonic paradigm is equivalent to assuming that
standard quantum mechanics is the absolute truth and everything else are
shadows.

In my opinion scientific theories, quantum mechanics in particular, are
something more modest. They are attempts at describing, rather imperfectly,
``the objective reality, which is independent of any theory''\cite{EPR}. In
consequence I prefer to retain as much as possible of the MWI, logically
superior to CI or MCI, but without adhering to the Platonic paradigm. The
choice is obvious to me: we should reject the completeness of quantum
mechanics, that rejection leading to the ensemble interpretation to be
commented in the next subsection. (But most people assume that MWI is not
compatible with an ensemble interpretation. Even if it is compatible the
relation is not trivial and will not be discussed here.)

We have seen that in CI and MCI, above commented, it is necessary to
introduce a probabilistic postulate, which is substituted for
Schr\"{o}dinger evolution equation during measurement. That postulate
(Born's rule) allows calculating the probabilities of the different possible
outcomes of a measurement and the corresponding state vector after the
collapse. In MWI it is controversial whether a probabilistic postulate is
introduced. Many authors consider that this is not the case, that the
quantum probabilities may be got from the formalism. Actually Everett
introduced, in his original formulation, a measure given by the squares of
the amplitudes in the sum (of normalized state vectors) which the world
state vector consists of. In our example, eq.$\left( \ref{4}\right) ,$ that
measure may allow to assume that $\left| c_{g}(t)\right| ^{2}$ and $\left|
c_{e}(t)\right| ^{2}$ are probabilities. Therefore it is my opinion that 
\emph{MWI does introduce a probabilistic postulate}, even if it is most
natural, as Everett emphasized\cite{Everett}.

An interesing consequence of the MWI is that a state vector is only
appropriate for the whole world. In contrast, the states of the systems
which we may actually study (subsystems of the universe) should be
represented by density matrices. This leads to the \emph{conjecture that
only a subset of the whole set of possible density matrices represent
physical states}. This conjecture strongly limits the validity of the
superposition principle and gives rise to the problem of determining what is
the subset of the whole set of density matrices which correspond to physical
(realizable) states. This problem will be discussed elsewhere.

\subsection{Ensemble interpretation and hidden variables}

The Copenhagen, von Neumann and many worlds interpretations have in common
the assumption that the description offered by quantum mechanics is
complete. They may be grouped within the class of `orthodox'
interpretations. An alternative to completeness is the assumption that the
wave function just represents our knowledge about the actual state of a
system. This hypothesis has been called the `ensemble interpretation' and it
was supported by Einstein\cite{EPR}, \cite{Einstein}, and also by some
authors in recent times\emph{\ }\cite{Ballentine}.

Th ensemble interpretation poses a question: What is the ensemble and what
is the probability distribution on the ensemble?. Answering the question
implies searching for an ontology behind the quantum formalism. That
research has been usually known as the \emph{hidden variables programme}.
The ensemble interpretation and the hidden variables approach are crucial
for a realistic understanding of quantum physics and consequently the whole
section 4 will be devoted to them.

\subsection{Conclusions}

In comparison with the Copenhagen and the von Neumann interpretations, the
many worlds (MWI) has the advantage that it does not require the measurement
postulates. It follows rigorously from the universal validity of quantum
mechanics. However in order to avoid a Platonic paradigm (see section 3.4),
strange to natural science, it should be combined with (or replaced by) an
ensemble interpretation, so giving rise to an interpretation which may be
realistic and not bizarre.

\section{The hypothesis that quantum theory is not complete}

\subsection{The epistemological versus ontological treatment of the
wavefunction}

Since the very early days of quantum mechanics the possibility was put
forward that the probabilistic character of the theory is due to the fact
that the description offered by the wave function is not complete. If this
is the case additional variables might be included in order to complete the
description. They do not appear explicitly in the quantum formalism whence
the name of `hidden variables'. However the mainstream of the scientific
community has been positioned against the hidden variables (HV). Possibly
the origin of this fact lies in the strong personality of Bohr, opposed to
HV, combined with the confort produced by the belief that one possesses the
final theoretical framework of physics, that is quantum mechanics. The
rejection to HV theories was reinforced by the failure to find a useful one.
In addition John von Neumann included in his celebrated 1932 book\cite{von
Neumann} a theorem apparently proving that any hidden variables model would
contradict the predictions of quantum mechanics. The theorem was an obstacle
for the research on the subject during more than three decades. In 1965 Bell%
\cite{Bell65} showed that the physical assumptions of von Neumann were too
restritive and that (contextual) hidden variables are possible\cite{Mermin}.
Indeed it is a simple matter to find a specific contextual hidden variables
model for any (simple) experiment consisting of a preparation followed by a
single measurement\cite{Scontext}. What is difficult is to get a HV model
valid for different experiments, for instance for a set of identical
preparations followed by several alternative measurements, as in typical
tests of Bell inequalities. Actually many of the supporters of the
incompleteness of the quantum description have not proposed a search for
specific HV models. This position would make the ensemble interpretation of
quantum mechanics a rather philosophical belief.

The dichotomy between completeness versus incompleteness of quantum
mechanics or, in modern language, epistemological versus ontological
treatment of the wavefunction, has been the subject of a controversy lasting
along the whole existence of quantum mechanics. As is well known, in the
early period the most famous debate took place between Bohr and Einstein
(see, e. g., \cite{Jammer} ). An important contribution to that debate was
the 1935 paper by Einstein, Podolsky and Rosen\cite{EPR}. Although the paper
is currently celebrated for having put forward the relevance of the
entanglement between distant particles, its declared purpose was to provide
(supposedly strong) arguments against the completeness of quantum mechanics.
In the paper the authors considered a system consisting on two particles
placed at a distance in a quantum state such that the particles are
correlated in both position and momentum. The state is possible according to
the quantum formalism and in fact the authors wrote explicitly the wave
function of the composite system. According to Heisenberg uncertainty
principle it is not possible to determine simultaneously the position and
the momentum of one of the particles but nothing forbids measuring only one
of the two observables with good accuracy. Due to the correlation, if the
position of one particle, say number 1, is measured we will know the
position of particle number 2 without interacting with it in any way. Thus,
after the measurement, we may attribute to the second particle a
wavefunction representing a state with definite position (but indefinite
momentum). On the other hand a measurement of the momentum of the first
particle allows attaching to the second particle a wavefunction
corresponding to a definite momentum (but indefinite position). The point of
the argument is that the state of the second particle should be the same in
both cases, because nothing has perturbed its state, and nevertheless we may
describe that state by means of two different wave functions. Hence the
authors concluded that the wave function just describes our knowledge and
not the real state of the particle. That is the wave function should be
treated as epistemological rather than ontological.

Crucial for the EPR\cite{EPR} argument is the assumption that \textit{no
influence could exist on a particle due to a measurement performed on
another distant particle, a hypothesis known as `locality'}. Bohr rebutted%
\cite{BohrEPR} the EPR argument claiming that in quantum mechanics there is
a kind of wholeness such that the assumed locality is not true. The current
wisdom, resting upon Bell's theorem to be discussed below, is that Bohr was
right and EPR were wrong.

In addition to nonlocality, a bizarre consequence of assuming an ontological
status for the wave function is exhibited in the EPR example. Indeed the
wave function of the two-particle system represents a pure state but the
state of each particle is not pure, according to the quantum rules. In fact
the state of the two-particle system is represented by a wave function, what
in standard quantum mechanics means that our knowledge of the two-particle
state is complete. However the state of one of the particles cannot be
represented by a wave function, but by a density operator obtained by taking
the partial trace of the density operator associated with the two-particle
wave function (the process is similar to the one leading to eq.$\left( \ref
{3a}\right) )$. That density operator represents a statisical mixture,
meaning incomplete knowledge. (I must point out that this fact does not
contradict the representation by a wave function made by EPR as commented
above. In the EPR argument the quantum state attributed to particle 2
follows from a measurement performed on particle 1, but now we are
considering the state when no measurement is made). The conclusion is that
we have complete information about a composite system, but incomplete about
every part, contrary to the usual definition of `complete'. It is as if a
student claims to know \emph{completely} the subject matter of a given book,
but she/he is admittedly \emph{ignorant about every chapter}. In my opinion
this behaviour of entangled quantum systems is another argument for the
epistemological character of the wavefunction. If that character is assumed,
our knowledge will be incomplete for both the composite (entangled) system
and every one of its parts, whence no paradox would arise.

In spite of the above arguments, during the whole history of quantum
mechanics the `orthodox view' has been that the theory is \emph{complete},
as stressed by Bohr and his followers. However Bohr's completeness may be
seen as a support to the `instrumentalistic approach' rather than a
statement about the relation between the wavefunction and reality. In
contrast for Einstein the relevant question was whether ``the $\psi
-function $ corresponds to a single system or to a (statistical) ensemble of
systems''\cite{Einstein} (Einstein carefully avoided the name wave function
-not to be commited himself to the existence of waves associated to
particles- and he used instead the name $\psi -function)$. He clearly
supported the latter assumption, which may be stated saying that he adhered
to the interpretation of the wave function as information. This
interpretation has been vindicated by recent authors, for instance Chris
Fuchs who has written ``quantum states are states of information, knowledge,
belief, pragmatic gambling commitments, not states of nature.''\cite{Fuchs}.
See also Englert and references therein\cite{Englert}\textrm{.}

At this moment it is appropriate to emphasize that an epistemolotical
interpretation of the wave function does not always imply for it a purely
subjective character. In many cases the available information is such that
everybody would attribute the same wave function to the physical system,
whence it acquires some objective character. A related question is whether
the wave function collapse after a measurement is a physical change or just
a change in our information (see section 3.3). In my opinion it is wrong to
adhere exclusively to one of the possibilities Actually both, or a
combination of both, may appear in measurements. The EPR argument provides
an example of a pure change of information, but an atom crossing a
Stern-Gerlach apparatus may suffer an actual physical change in the
direction of the spin. In any case the quantum postulate that `in a
measurement the state of the system goes to an eigenstate of the measured
observable' may be appraised as an elegant formal statement, but in actual
experiments things are more involved.

\subsection{Recent approach to realistic intepretations}

The advances in quantum information theory during the last three decades
have had an important influence on foundations. For our purposes three
aspects closely related are relevant: 1) A renewed support to the assumption
that the quantum wave function (or state vector) represents information, 2)
A vindication of some Einstein's views on the interpretation of quantum
mechanics, and 3) A study of the foundational problems from the point of
view of realism. The recent vindication of Einstein does not refer to his
opinions (it is generally assumed that he was wrong in his beliefs on
locality, allegedly refuted by Bell's theorem, but see section 5.) Rather he
is vindicated as having pointed out what are the relevant questions to be
answered. In fact a close scrutiny of Einstein's letters to different
authors shows that his main interest was not the question whether the
wavefunction $\psi $ represents an ensemble of possible systems - or, what
is almost equivalent, if it only represents our information- but on whether
a given real (ontic) state may correspond to different quantum-mechanical
states $\psi .$ Einstein clarified the point in a letter to Schr\"{o}dinger%
\cite{Lewis}. An extended discussion about Einstein's opinions, with many
references, appears in a paper by Harrigan and Spekkens\cite{Spekkens07}.

In recent times it has become popular to study the possibility of realistic
interpretations of quantum theory resting upon the concept of `ontic
states', that is real physical states of systems not necessarily completely
described by quantum theory, but objective and independent of the observer%
\cite{Spekkens07}. In the following I will use the names `hidden variables
models' or `ontological models' as equivalent.

Thus the standard assumption in classical physics that measurements just
reveal existing properties may be formalized stating that the observed
results depend on both the ontic state, $\lambda ,$ of the system and the
measuring apparatus, $A,$ appropriate for a given observable quantity. That
is the observed result, $a$, will be a function 
\[
a=a\left( \lambda ,A\right) . 
\]
Let us analyze whether a similar analysis can be made in quantum physics. If
we assume that the results of all observations on a system derive from
functions like $a\left( \lambda ,A\right) ,$ then the correlation between
several observable properties, $\left\{ A,B,...C\right\} ,$ may be written 
\begin{equation}
\left\langle AB...C\right\rangle =\int f_{\psi }\left( \lambda \right)
a\left( \lambda ,A\right) b\left( \lambda ,B\right) ...c\left( \lambda
,C\right) d\lambda ,  \label{Bellhv}
\end{equation}
where $f_{\psi }\left( \lambda \right) d\lambda $ gives the probability
distribution of the ontic states in a given quantum state, either pure or
mixed (but we use the subindex $\psi $ for both cases). Without loss of
generality we may consider that the properties $\left\{ A,B,...C\right\} $
can take only the values $\left\{ 0,1\right\} $ because any property may be
defined in terms of yes-no questions. Thus the knowledge of all correlations
like eqs.$\left( \ref{Bellhv}\right) $ determines the joint probability
distribution of all observable properties and the reciprocal is also true.
We may define noncontextual hidden variables models (HVM) (or noncontextual
ontological models) as those where all correlations may be got from eqs.$%
\left( \ref{Bellhv}\right) $. The predictions of quantum mechanics not
always can be interpreted that way. In fact the Kochen-Specker theorem
proves that noncontextual HVM are not always possible.

The correlations involved in eq.$\left( \ref{Bellhv}\right) $ may appear in
two different scenarios: 1) Correlations between properties of a system
localized in a small region of space, 2) Correlations between distant
systems. Actually the difference between the two scenarios is not sharp, but
there is an important case which belongs clearly to the latter class, namely
in EPR type experiments to be discussed below, when dealing with Bell's
theorem.

Recently a theorem has been proved by Pusey, Barrett and Rudolph\cite{Pusey}
which apparently implies that quantum states are physical properties of a
system, contrary to Einstein's opinion. If this implication is correct, the
theorem would be very important, because it would contradict the assumption
that the wave function represents only information and this is the
cornerstone for the realistic interpretation supported in this paper. The
authors claim to have proven that, for two different quantum states
represented by the wavefunctions $\psi $ and $\phi ,$ ``the distributions $%
f_{\psi }\left( \lambda \right) $ and $f_{\phi }\left( \lambda \right) $ (of
ontic states $\lambda $) cannot overlap. If the same can be shown for any
pair of quantum states, then the quantum state can be inferred uniquely from 
$\lambda $. In this case, the quantum state is a physical property of the
system''.

In order to see the relevance of the theorem for a realistic interpretation
let us return to the example of the Schr\"{o}dinger cat. In standard quantum
mechanics it is assumed that any system is in a quantum state (represented
by a wave function), although the most useful representation for macroscopic
bodies is a density operator (equivalent to a probability distribution of
state vectors or wave functions). Thus our living cat will be in some
quantum state represented by one of the state vectors $\mid livecat,j\rangle
,$ j = 1,2,... Similarly a dead cat may be represented by $\mid
deadcat,k\rangle ,$ k= 1,2,... But standard quantum mechanics also assumes
that a linear combination like 
\begin{equation}
\frac{1}{\sqrt{2}}(\mid livecat,1\rangle +\mid deadcat,1\rangle ),
\label{Pusey}
\end{equation}
also represents a possible quantum state. Now the commented theorem\cite
{Pusey} implies that all ontic states associted to the quantum state eq.$%
\left( \ref{Pusey}\right) $ should be different from every ontic state of
living cat, associated to the quantum state$\mid livecat,j\rangle ,$ and
also from every ontic state of dead cat, associated to $\mid
deadcat,k\rangle .$ But no plausible realistic interpretation may assume the
existence of ontic states associated specifically to the quantum state eq.$%
\left( \ref{Pusey}\right) $ (of partially living cat!) Any realistic
interpretation of that quantum state should associate to it a statistical
mixture of ontic states of living cat and dead cat.

Thus a careful scrutiny of the assumptions of the theorem is necessary.
According to the authors the assumptions are: ``a system has a real physical
(ontic) state... This assumption only needs to hold for systems that are
isolated, and not entangled with other systems... The other main assumption
is that systems that are prepared independently have independent physical
states''\cite{Pusey}. However asides from the explicit hypotheses there are
implicit assumptions, for instance that linear combinations like eq.$\left( 
\ref{Pusey}\right) $ represent possible quantum states.\ A critical analysis
of all these implicit assumptions lies beyond the scope ot this paper.

\subsection{Bell\'{}s theorem}

The original theorem of Bell\cite{Bell}, \cite{Bell87}, \cite{Mermin}
provides necessary conditions for the possibility that measurements
performed in two distant regions are independent, i. e. they cannot
influence each other. More formally Bell\'{}s fundamental hypothesis may be
stated as follows. Let us assume that pairs of particles (more generally
subsystems) are produced in a source and the two particles of every pair
move in different directions. Eventually one of the particles arrives at
Alice, who measures some observable $A$, and the other one arrives at Bob,
who measures the observable $B$. If the result obtained by Alice (Bob) in
one run of the experiment is $a_{j}$ ($b_{j})$ , after a large number, $n$,
of similar runs the relevant quantity is the correlation 
\begin{equation}
\left\langle AB\right\rangle _{n}\equiv \frac{1}{n}\sum_{j=1}^{n}a_{j}b_{j}.
\label{AB}
\end{equation}
Here `similar' means that the runs of the experiments, each consisting of
the preparation of a pair and the subsequent measurements, are performed in
identical conditions, as far as they may be controlled (we cannot exclude
that perturbations out of control may arise in every run.) Bell assumed that
the result of Alice's measurement depends on the values of the (hidden)
variables, collectively labelled $\lambda _{a},$ that specify the real state
(the `ontic' state) of the Alice\'{}s particle and, obviously, on the
Alice's measuring set up, which will be labelled $A$ here. We shall write
that dependence in the form of a function $a\left( \lambda _{a},A\right) $.
Similarly for Bob's particle we write the function $b\left( \lambda
_{b},B\right) .$ Thus the theoretical correlation, see eq.$\left( \ref{AB}%
\right) ,$ may be written 
\begin{equation}
\left\langle AB\right\rangle =\int \rho \left( \lambda _{a},\lambda
_{b}\right) a\left( \lambda _{a},A\right) b\left( \lambda _{b},B\right)
d\lambda _{a}d\lambda _{b},  \label{BellAB}
\end{equation}
where $\rho \left( \lambda _{a},\lambda _{b}\right) $ is the joint
probability density for the variables $\lambda _{a}$ and $\lambda _{b}$
(compare with eq.$\left( \ref{Bellhv}\right) ).$ The essential assumption of 
\emph{locality} is that neither $a$ depends on $B$ nor $b$ depends on $A$,
nor $\rho $ depends on either $A$ or $B$. As usual, the probabilities
involved in eq.$\left( \ref{BellAB}\right) $ are tested by measuring the
frequencies appearing in eq.$\left( \ref{AB}\right) $ with $n$ large.
Actually we might include in the functions $a\left( b\right) $ additional
hidden variables $\mu _{a}$ ($\mu _{b})$ taking into account the fact that
no measuring set up may be completely controlled. After that we should
perform appropriate averages over those variables, which is equivalent to
assuming that $a\left( \lambda _{a},A\right) $ and $b\left( \lambda
_{b},B\right) $, eq.$\left( \ref{BellAB}\right) ,$ are already averages over
the variables $\mu _{a}$ and $\mu _{b},$ respectively.

From Bell's proposal, eq.$\left( \ref{BellAB}\right) ,$ it is a trivial task
to derive inequalities that are necessary conditions for the existence of
local models of the correlation experiments. Bell also proved that there are
(ideal) experiments where quantum mechanics predicts violations of one of
the inequalities. Hence Bell's theorem follows: ``Local hidden variables
models of quantum mechanics are not possible''.

Actually there is an important consequence of Bell's work that is
independent of the existence of quantum mechanics, and not always has been
duly appreciated. That is, \emph{if there are correlations between distant
systems which violate a Bell inequality then those correlations cannot be
explained as deriving from a common cause.} In fact any such explanation
could be formalizd by eq.$\left( \ref{BellAB}\right) $ and it would imply a
Bell inequality. The relevance of this result is that the explanation of
correlations between distant bodies as deriving from a common past is a
cherised hypothesis, not only in physics but in all branches of science and
even in ordinary life. For instance everybody will believe that the
similarity between twins is due to the common genetic code.

For many quantum physicists, not too fond of foundations, the merit of
Bell's theorem was to ``show the absurdity of searching for hidden variables
of quantum mechanics, an useless goal in the first place.'' However the
relevance of Bell\'{}s theorem is greater than just to refute a class of
hidden variables theories of quantum mechanics, as pointed out above.
Furthermore the theorem implies that there is some conflict between
relativity theory and quantum mechanics. Indeed Bell himself reinterpreted
eq.$\left( \ref{BellAB}\right) $ in the sense that $\lambda _{a}$ and $%
\lambda _{b}$ mean the set of all events in the past light cones of the
measurements performed by Alice and Bob, respectively. Thus if the
measurements are space-like separated, in the sense of relativity theory,
then Bell's theorem seems to prove the incompatibility of quantum mechanics
with relativity theory. The contradiction is dramatic, but people have found
an scape after the proof that experiments violating a Bell inequality do not
allow sending superluminal signals from Alice to Bob (or from Bob to Alice)
and this is the only think forbidden by relativity theory. In my opinion
this is not a satisfactory solution. I am convinced that we should hold
strong on the validity of the principles of realism and (relativistic)
locality in physics. This was also the belief of Einstein until his death%
\cite{Einstein}.

Quantum mechanics has had so spectacular a success in predicting the results
of experiments that for most authors any proposal to change the quantum
formalism seems a blunder. But I am convinced that \emph{a solution must be
found} to the conflict posed by Bell's theorem. Thus I propose the following 
\emph{restriction - }not a modification - of the quantum formalism.

\begin{conjecture}
Experiments showing a (loophole-free) violation of a Bell inequality are not
feasible.
\end{conjecture}

To many readers this conjecture may appear as a pure speculation. But it is
a scientific statement in Popper's sense because it can be tested, and
eventually refuted, by experiments. On the other hand the conjecture will be
increasingly confirmed as time passes without a refutation, and half a
century has already elapsed from Bell's work\cite{Kwiat}. It is true that
the experiments have been refined along the time and that quantum mechanics
has been vindicated in those experiments, with a few exceptions not too
significant. In any case we should conclude that the question is open. In
the following an attempt will be made to convince readers that the above
conjecture is not crazy.

Firstly it is necessary to point out a fact usually neglected. The most
spectacular agreement between the quantum predictions and the experiments
may be explained from the quantum equations and a little more. E.g. the
calculation of the anomalous magnetic moment of the electron, in agreement
with experiments up to 10 decimal figures, derives from the solution of the
quantized coupled Dirac and Maxwell equations for an electron placed in a
homogeneous magnetic field. Asides from the quantized Dirac and Maxwell
equations no additional \emph{quantum assumptions} are needed. In particular
we may dispense with the quantum postulates of measurement because in the
experiment the light detection may be described as a macroscopic process. 
\emph{The precision of the agreement between theory and experiment compel us
to admit that the quantum equations are correct.} In contrast the
measurement of probabilities rarely provides an agreement better than a few
percent. For instance in performed tests of Bell inequalities the measured
parameter, that is a linear combination of probabilities, is typically
reported with uncertainty greater than one per thousand\cite{Kwiat}.

Secondly it is physically absurd, although may appear as mathematically
elegant, to assume an one-to-one correspondence between density matrices and
states of a system (or between `positive operator valued measures', POVM,
and feasible measurements). In particular only a relatively small number of
different states of a system may be actually prepared in the laboratory
whilst the set of quantum state vectors consists of infinitely many, and
similarly for the measurements. Most textbooks are aware of the problem and
present the quantum postulates without assuming one-to-one correspondence.
It is stated, for instance, that ``to every state of a physical system we
associate a density matrix'', but without postulating the reciprocal.
Nevertheless in the proof of Bell's theorem a quantum state vector violating
a Bell inequality is used without investigating whether the corresponding
state may be actually produced in the laboratory. In other words Bell's, as
any theorem, is a \emph{mathematical }statement, a contradiction between the
quantum formalism and the assumption eq.$\left( \ref{BellAB}\right) $. The
true, very important, physical consequence of Bell's work is to \emph{%
suggest experiments} that may eventually refute local realism, but directly
the theorem proves nothing about nature.

Thirdly quantum theory itself puts some constraints on the feasibility of
experiments, difficulties usually dismissed as (small) practical
defficiencies. For instance in recent tests involving entangled photons
produced in a nonlinear crystal an efficient photon detection cannot be
achieved using too short a time window, which puts a challenge for the
spacelike separated detections needed to close the locality loophole\cite
{Kwiat}.

\section{Proposed physical model of the quantum world}

A realistic interpretation, giving rise to an ontology, that is a physical
model of the world, would make quantum mechanics more palatable to lovers of 
\emph{theory}, in the ethimological sense of contemplation. It would allow
``understanding quantum mechanics''\cite{Feynman}. In the model here
proposed I shall not consider the quantization of gravity because I think
that only after having a good understanding of quantum mechanics in Minkowsi
space might we try to understand quantum gravity. The picture here supported
was sketched in the Introduction section and further discussed in the
following, where the most relevant assumptions will be presented as
`propositions'.

In spacetime there are fields, a form of `matter' giving rise to phenomena
that may be observed. In quantum theory every field at a spacetime point is
represented by a (scalar, vector or tensor) operator belonging to some
non-commutative algebra (whose full mathematical structure is well known and
will not be specified here). I shall refrain from making any hypothesis
about the state of the universe as a whole. I will consider only the study
of systems of `human size' (rather than cosmic size). For instance an atom
or a small piece of bulk matter.

The crucial assumption is that the representation of fields by operators,
rather than classical functions, should be interpreted as the fact that the
fields are stochastic. That is, I assume the following

\begin{proposition}
The quantum formalism is a disguised form of dealing with \emph{peculiar
stochastic fields}. Furthermore the quantum vacuum fluctuations of all those
stochastic fields are real. Commutation or anticommutation rules of the
field operators characterize the stochastic properties of the fields.
\end{proposition}

The assumption puts a well known and strong problem, namely that the free
fields are ultraviolet divergent. Solving the problem lies outside the scope
of this paper. As possible solutions I suggest either some kind of
cancelation for interacting fields or general relativistic effects. An
important consequence of the proposition is

\begin{proposition}
No small system can be isolated from the rest of the world, even
approximately.
\end{proposition}

Indeed every system should be effectively interacting with many other
systems via the vacuum fields. But in order to be able to make physics we
should assume that microscopic systems, even if not isolated, may be treated
with a formalism that in some form takes into account the interaction. I
believe that this formalism is quantum mechanics. For instance, if we
represent the state of an atom by a state vector it is plausible to assume
that this representation corresponds to the atom `dressed' by all fields
that interact with it. This is consistent with the fact that in quantum
electrodynamics the physical electrons are never `bare' but `dressed with
virtual photons and electron-positron pairs'. The word `virtual' is just a
name for something that we know to have observable effects, but we cannot
consider `real' without a conflict with the cherised (but for me wrong)
assumption that quantum systems may exist as isolated. The point is that the
representation of an atom by a state vector takes into account the
(approximate) action of the vacuum fields as is shown by the use of the
physical, rather than bare, mass and charge of the electrons. And similarly
for other quantum systems.

As a consequence it is a rather presumptuous attitude to pretend that a
state vector represents faithfully the \emph{actual state of an individual
system}. It is more plausible to assume that the state vector represents the
relevant information available about the system. The conclusion is that the
quantum-mechanical representation of the state of a system is incomplete.
This incompleteness is the cause of the claimed `irreducible probabilistic
character of the physical laws''. For instance the fact that an atom decays
at a time that cannot be predicted derives from the fluctuations of the
vacuum fields that actually stimulate the decay.

The concept of isolated system is the cornerstone of classical physics and,
therefore, it is not strange that it was also introduced in quantum physics.
It is true that early authors dealing with quantum theory, like Planck,
Einstein and Nernst, studied the possible existence and influence of vacuum
(nonthermal) fluctuations\cite{Milonni}. However the success of the Bohr
atomic model, where the concept of fluctuation was absent, reinforced the
idea that quantum systems may be treated as isolated. Nonthermal
fluctuations reappeared in modern quantum mechanics associated to the zero
point energy of bounded quantum systems. However in the alternative of
either rejecting the assumption of isolated system or dismissing the reality
of the quantum fluctuations, the mainstream of the community choosed the
latter. This compelled people to introduce the ill-defined concept of
`virtual'. In my opinion that choice has been the source of most
difficulties for a realistic interpretation of the quantum formalism.

The existence of real vacuum fluctuations gives rise to two characteristic
traits of quantum physics. Firstly quantum theory should be \emph{%
probabilistic}. Secondly it should present a kind of \emph{wholeness}, quite
strange to classical physics where the concept of isolated system is
crucial. The fact that the vacuum fluctuations at different points may be 
\emph{correlated} is the origin of the wholeness, which manifests specially
in the phenomenon of \emph{\ entanglement.}

\begin{proposition}
In addition to the usual correlations between (two or more) physical
systems, involving directly observable quantities, there are additional
correlations via the quantum vacuum fluctuations interacting with the
systems.\emph{\ }These correlations give rise to the phenomenon of
entanglement.
\end{proposition}

This leads to the following \emph{picture of the quantum world}. Fundamental
fermions, like leptons or quarks, are (localized) particles, but fundamental
bosons, in particular photons, are actually (extended) fields. Gravity plays
a special role, I support the view that general relativity determines the
structure of (curved) spacetime and its relation with matter, so that
gravity is not a field in the same sense than other fields. The wave
behaviour of particles derives from the unavoidable interaction with fields,
and the particle behaviour of fields derives from the interaction with
particles, e. g. during detection. A fundamental property of the universe is
the existence of fluctuations of all fields in the vacuuum (i. e. at zero
Kelvin.) In the case of Bose fields there are random fluctuations similar to
the zeropoint fluctuations of the electromagnetic field. In the case of
Fermions the fluctuations may correspond to the existence of a kind of Dirac
sea of particles and antiparticles that may be created and annihilated.
There should be also metric fluctuations of spacetime itself, possibly
stronger than those deriving from the fluctuations of stress-energy of the
particles and fields. That is a background of gravitational waves with
wavelengths of atomic or subatomic size.

Now the question is to what extent the quantum formalism is compatible with,
or better encodes, this picture. It is necessary to distinguish the core
from the rest of the quantum formalism. The core consists of the Hilbert
space (or $C^{*}$-algebra) mathematical structure and the fundamental
equations of motion (Dirac, Maxwell, Klein-Gordon, etc.) in terms of that
structure. To the mathematical theory and the equations it is necessary to
add Born's rule. In the non-relativistic approximation to particle motion,
Born\'{}s rule is the interpretation of the modulus squared of the wave
function as a probability density. It should be appropriately generalized
for fields.

\emph{I do not propose any modification of that core, but claim that the
rest of the quantum formalism is dispensable}, although it may be useful in
practice. I think that it is flawed to introduce physical operations as a
part of the postulates of quantum mechanics, for instance `preparation' or
`measurement'. Therefore it is not appropriate to stablish any rigid
correspondence between `preparation' and `density matrix' or between
`measurement' and `selfadjoint operator' (or POVM). Preparation is a rather
complex set of physical manipulations whence the density matrix appropriate
for a microscopic system may be guessed, rather than derived, most times
after a process of trial and error on the part of the scientists performing
the particular preparation. Similarly for the measurement. Indeed an
empirical result is taken as a valid discovery only after the relevant
experiment has been critically analyzed and repeated by different groups of
researchers. This view agrees with what Bell wrote: ``I am now convinced
that the word `measurement' has now been so abused that the field would be
significantley advanced by banning its use altogether, in favour for example
of the word `experiment'.'' (\cite{Bell87}, page 166)

As a consequence I believe that a formalization of preparations and
measurements or theorems derived from it, like the one by Pusey et al.\cite
{Pusey} are of limited value. Furthermore, as pointed out in the subsection
devoted to Bell's theorem, it is an undue extrapolation to assume that all
density matrices may correspond to physical states. Hence it follows that
the boundary of the set of (physical) density opertors is not the set of
idempotent oparators (or what is equivalent, the set of state vectors) and
the very concept of `quantum pure state' is not well defined.

In summary I propose to search for a realistic interpretation of quantum
mechanics, or equivalently a physical model of the microworld, using only
the core of the formalism as defined above. That is the Hilbert space
structure, the equations and Born\'{}s rule, without attempting to attach a
meaning to the remaining postulates, which does not preclude their practical
usefulness.

\end{document}